\documentclass[aps, pre, english, showpacs, superscriptaddress, reprint]{revtex4-1}
\usepackage{graphicx}
\usepackage{color}
\usepackage{amssymb}
\usepackage{amsmath}
\usepackage{tabularx}
\usepackage{bm}
\usepackage{hyperref}
\usepackage{ltxtable}
\usepackage{natbib} 
\usepackage{longtable}
\usepackage{float}
\newcommand{\be}{\begin{equation}}
\newcommand{\ee}{\end{equation}} 
\newcommand{\bea}{\begin{eqnarray}}
\newcommand{\eea}{\end{eqnarray}}

\usepackage{array}

\usepackage[usenames,dvipsnames]{xcolor}

\usepackage[normalem]{ulem}
\newcolumntype{P}[1]{>{\raggedright\arraybackslash\color{blue}}p{#1}}

\hypersetup{
colorlinks,
citecolor=blue,
filecolor=blue,
linkcolor=blue,
urlcolor=blue}

\begin{document}
\title{Energy and enstrophy  spectra and fluxes for the inertial-dissipation range of  two-dimensional turbulence}

\author{Akanksha Gupta}
\email{akgupt@iitk.ac.in}
\affiliation{Department of Physics, Indian Institute of Technology, Kanpur 208016, India}
\author{Rohith Jayaram}
\affiliation{Department of Energy and Process Engineering, Norwegian University of Science and Technology,  Trondheim NO-7491, Norway}
\author{Anando G. Chaterjee}
\author{Shubhadeep Sadhukhan}
\affiliation{Department of Physics, Indian Institute of Technology, Kanpur 208016, India}
\author{Ravi Samtaney}
\affiliation{Mechanical Engineering, Division of Physical Science and Engineering, King Abdullah University of Science and Technology - Thuwal
23955-6900, Kingdom of Saudi Arabia}
\author{Mahendra K. Verma}
\email{mkv@iitk.ac.in}
\affiliation{Department of Physics, Indian Institute of Technology, Kanpur 208016, India}
\date{\today}

\begin{abstract}
In this paper, using Pao's conjecture  [Y.-H. Pao, Phys. Fluids 8, 1063 (1965)],  we derive expressions for the spectra and fluxes of kinetic energy and enstrophy for two-dimensional (2D) forced turbulence  that extend beyond the inertial range.  In these expressions, the fluxes and the spectra contain additional factors of the exponential form.  To validate these model predictions, we perform numerical simulations of 2D turbulence with external force applied at $k=k_f$ in the intermediate range.  The numerical results  match with the model predictions, except for the energy and enstrophy fluxes for  $k<k_f$, where the fluxes exhibit significant fluctuations.   We show that these fluctuations arise due to the unsteady nature of the flow at small wavenumbers.   For the $k<k_f$, the shell-to-shell energy transfers computed using numerical data show forward energy transfers among the neighbouring shells, but backward energy transfers for other shells.  
\end{abstract}
\maketitle

\section{Introduction} \label{sec:Introduction}
Turbulence is an omnipresent phenomena~\cite{McComb:book:Turbulence,Frisch:book,Spiegel:book_edited:Turbulence,Lesieur:book:Turbulence,Alexakis:PR2018}.  Though many natural and  laboratory flows are three-dimensional, many  astrophysical and geophysical flows exhibit two-dimensional (2D) or quasi two-dimensional  behavior~\cite{Kolesnikov:FD1974,Kellay:RPP2002,Tabeling:PR2002,Clercx:AMR2009,Verma:EPL2012,Boffetta:ARFM2012}. For example, strong rotation suppresses the velocity component in  the direction of rotation~\cite{Oks:POF2017,Xia:POF2017,Sharma:PF2018}.  Similarly, a strong external magnetic field in magnetohydrodynamics~\cite{Potherat:JFM2000,Lee:ApJ2003,Reddy:PF2014,Verma:ROPP2017}, and  strong gravity in planetary environments~\cite{Lindborg:PF2010,Davidson:book:TurbulenceRotating,Verma:book:BDF}  make the flow quasi two-dimensional.   Therefore, a good understanding of 2D  turbulence  is important for modeling such flows.  In this paper, we address the  spectra and  fluxes of energy and enstrophy for the inertial-dissipation range of 2D turbulence. 

Using analytical arguments, \citet{Kraichnan:PF1967_2D}   predicted a dual cascade  for 2D  turbulence that is forced at an intermediate scale ($k \approx k_f$).  He showed an inverse cascade of kinetic energy for $k<k_f$, and forward cascade of enstrophy for $k>k_f$.   In the corresponding regimes, the kinetic energy spectra are $E_u(k)=C\epsilon^{2/3}k^{-5/3}$ and $E_u(k)=C' \epsilon_\omega^{2/3}k^{-3}$ respectively; here $\epsilon$, $\epsilon_\omega$ are respectively the energy and enstrophy dissipation rates (or injection rates), and $C,C'$ are  constants.     Numerical simulations and analytical calculations indicate that $C \approx 6.5 \pm 1$, and $C' \approx 1.0 \pm 1$~\cite{Boffetta:ARFM2012,Gotoh:PRE1998}.  Further,  \citet{Kraichnan:JFM1971_2D3D} derived a logarithmic correction to the latter spectrum.    Using the properties of structure function,  \citet{Gotoh:PRE1998} generalised the spectrum of forward enstrophy cascade regime to the dissipation range.    He argued that $E_u(k) \propto  k^{-(3+\delta)}$ for $k<k_{d2D}$, and $E_u(k) \propto  k^{-(3+\delta)/2} \exp(-\alpha_2 k/k_{d2D})$ for $k>k_{d2D}$, where $k_{d2D} = \epsilon_\omega^{1/6} /\sqrt{\nu}$ is the enstrophy dissipation wavenumber, and $\delta, \alpha_2$ are constants.  \citet{Gotoh:PRE1998} also verified the above scaling using numerical simulation.

Kraichnan's formulas for the dual energy spectrum have been observed in many laboratory experiments, for example by Paret and
Tabeling~\cite{Paret:PRL1997}, Rutgers~\cite{Rutgers:PRL1998}, and Kellay et al.~\cite{Xiong:EPL2011}. In numerical simulations, the same  phenomena has also been observed by Siggia and Aref~\cite{Siggia:PF1981}, Frisch and
Sulem~\cite{Frisch:PF1984}, and Borue~\cite{Borue:1994}.   For a forced  2D turbulence,  the large-scale energy  grows in
time~\cite{Smith:PRL1993}.  In the regime with forward enstrophy transfer, the energy spectrum is typically steeper than $k^{-3}$, both in numerical simulations
~\cite{Legras:1988} and in experiments~\cite{Kellay:PRL1995}.  Moreover, \citet{scott:2007,Fontane:2013} report some deviations from the theoretical predictions of \citet{Kraichnan:PF1967_2D}.  \citet{Pandit:2017bi} describe properties of 2D flows in the presence of complex forces. \citet{Masih:JOAS2018} studied the energy transfer between the synoptic scale and the mesoscale using DNS of 2D turbulence under forcing applied at different scales.

\citet{Boffetta:JFM2007} performed direct numerical simulations of forced 2D Navier-Stokes equations and studied  the energy and enstrophy cascade regimes with good accuracy. \citet{Boffetta:JFM2007} employed Ekman friction to suppress  energy growth at large scales.  Besides the above spectral laws,  variable energy flux, irregular and non-local energy transfer have also been studied for 2D turbulence~\cite{Danilov:2001}.  \citet{Musacchio:PRF2019} investigated the formation of large-scale structures in a turbulent fluid confined in a thin layer. However, despite many years of work, there are some discrepancies on the scaling laws.  Also, see \citet{Alexakis:PR2018} for description of various properties of energy fluxes, including those of 2D turbulence.

Kolmogorov's theory~\cite{Kolmogorov:DANS1941Structure,Kolmogorov:DANS1941Dissipation} yields $k^{-5/3}$ energy spectrum for 3D hydrodynamic turbulence.  \citet{Pao:PF1965,Pao:PF1968}  generalised this scaling to inertial-dissipation range by postulating that the ratio of the energy spectrum and energy flux is independent of the kinematic viscosity, and that it depends on the dissipation rate and local wavenumber. We employ Pao's conjecture \cite{Pao:PF1968}  to 2D turbulence, and extend the  $k^{-5/3}$ and $k^{-3}$ spectra and corresponding fluxes  \citep{Kraichnan:PF1967_2D}  beyond the inertial range.  

We  simulate 2D turbulence numerically  and compute the spectra and fluxes of energy and enstrophy, and compare the numerical results with the predictions of extended model of spectra and fluxes based on Pao's conjecture.  We observe  good agreement between the numerical and model results for $k>k_f$.  However, they differ for $k<k_f$ possibly due to the unsteady nature of 2D turbulence.   

The present paper is structured as follows. In  Sec.~\ref{sec-goveq}, we describe the governing equations for a forced two-dimensional incompressible fluid. In  Sec.~\ref{sec:theory}  we derive the  spectra and fluxes of energy and enstrophy using Pao's conjecture. In Sec.~\ref{Sec:simdetails}, we describe our numerical procedure and parameter values.  Sec.~\ref{Sec:results} contains simulation results and comparison with model predictions. We conclude  in Sec.~\ref{Sec:conclusion}.

\section{Governing Equations}
\label{sec-goveq}
The Navier-Stokes equations for a forced two-dimensional incompressible fluid is
\begin{eqnarray}
\frac{\partial \mathbf{u}}{\partial t} + \mathbf{u}\cdot \nabla \mathbf{u} &=& -\nabla p +\nu \nabla^2 \mathbf{u}+ \mathbf{F_{u}}, \label{NSE}\\
\nabla \cdot \mathbf{u} &=&  0, \label{incompress}
\end{eqnarray}
where $\mathbf{u}$ and $p$ are the velocity and pressure fields respectively, $\nu$ is the kinematic viscosity, and $\mathbf{F_{u}}$ is the external force. We take density to be constant ($\rho=1$). The flow is two-dimensional in $xy$ plane, and the vorticity is a scalar: $\omega =(\nabla  \times \mathbf{u}) \cdot \hat{z}$. Taking a curl of Eq.~(\ref{NSE}) yields the following  dynamical equation for the vorticity field:
\begin{eqnarray}
\frac{\partial \omega}{\partial t} + \mathbf{u}\cdot \nabla \omega &=& \nu \nabla^2 \omega+ F_{\omega},
\label{NSEV}
\end{eqnarray}
where $F_{\omega} = [\nabla \times \mathbf{F_{u}}]_z$. 

For  a 2D hydrodynamic flow, the  total kinetic energy (KE), $E_u$, and  the total enstrophy, $E_\omega$, are  defined below:
\begin{equation}
E_u = \int d{\bf r} u^{2}({\bf r})/2;~~~E_\omega = \int d{\bf r} \omega^2({\bf r})/2
\end{equation}
These quantities are conserved for 2D flows in the inviscid and force-free regime~~\cite{McComb:book:Turbulence,Frisch:book,Spiegel:book_edited:Turbulence,Lesieur:book:Turbulence}.  These quadratic invariants play an important role in 2D turbulence.

The above equations for the velocity and vorticity fields are written in Fourier space as
\begin{eqnarray}
  \frac{d}{d t}  {\bf u} (\mathbf{k}) + {\bf N}_u ({\bf k})
 &=& -i {\bf k} p (\mathbf{k}) + {\bf F}_u({\bf k})  -  \nu k^{2}  {\bf u}(\mathbf{k}), \label{eq:fourier:uk}\\
  \frac{d}{d t}  {\omega}(\mathbf{k}) + N_\omega ({\bf k}) &=&  F_\omega(\mathbf{k}) -  \nu k^{2} {\omega}(\mathbf{k}) , \label{eq:fourier:omegak} \\
    {\bf k\cdot u} (\mathbf{k})  &=&  0, \label{eq:fourier:k_uk_zero}
\end{eqnarray}
where
\begin{eqnarray}
{\bf N}_u ({\bf k})  &=& i  \sum_{\bf p}  {\bf \{ k \cdot u(q) \} u(p) }, \label{eq:fourier:Nuk} \\
N_\omega ({\bf k}) &=& i  \sum_{\bf p}  {\bf \{ k \cdot u(q) \} \omega(p) }, 
  \label{eq:fourier:Nomega1}
\end{eqnarray}
with ${\bf q=k-p}$.    Note that the pressure $p (\mathbf{k})$ is derived by   taking dot product of  Eq.~(\ref{eq:fourier:uk}) with $i {\bf k}$ and by employing ${\bf k\cdot u}({\bf k})=0$:
\begin{eqnarray}
 p(\mathbf{k}) &=&  \frac{i}{k^2}  {\bf k} \cdot \{ {\bf N}_u({\bf k}) - {\bf F}_u({\bf k}) \}.
 \label{eq:fourier:Pk}
\end{eqnarray}

 To derive a dynamical equation for the modal KE $E_u({\bf k}) = |{\bf u}({\bf k})|^2 /2$ and modal enstrophy  $E_\omega({\bf k})  = |{\omega}({\bf k})|^2 /2$, we perform dot products of Eq.~(\ref{eq:fourier:uk}) with ${\bf u}^*({\bf k})$, and Eq.~(\ref{eq:fourier:omegak}) with with $\omega^*({\bf k})$, and add the resultant equations with their complex conjugates. These operations yield  
\begin{eqnarray}
 \frac{d}{dt} E_u(\mathbf{k}) & = &   \sum_{\bf p} \Im \left[ {\bf \{  k \cdot u(q) \}  \{ u (p) \cdot  u^*(k) \} }  \right] \nonumber \\
&& + \Re[ {\bf F}_u({\bf k}) \cdot  {\bf u^*(k)}] - 2 \nu k^{2} E_u({\mathbf k}) 
 \nonumber \\
 & = & T_u({\bf k}) + \mathcal{F}_u({\bf k})- D_u ({\bf k})\label{eq:fourier:Euk} 
 \end{eqnarray}
 and
 \begin{eqnarray}
\frac{d}{dt} E_\omega(\mathbf{k})  & = &    \sum_{\bf p} \Im \left[ {\bf \{ k \cdot u(q) \}  \{  \omega(p)  \omega^* (k) \} } \right] \nonumber\\
 &&+\Re[ F_\omega({\bf k})   {\omega}^*({\bf k}) ]) - 2 \nu k^{2} E_\omega({\mathbf k}) \nonumber \\
 & = & T_\omega({\bf k}) + \mathcal{F}_\omega({\bf k})- D_\omega ({\bf k}),
 \label{eq:fourier:Eomega_k_dot_2D}
\end{eqnarray}
where  $\Re[ . ]$, $\Im[ . ]$ are  real and imaginary parts of the argument respectively;  $T_u({\bf k}), T_\omega({\bf k})$  are respectively the rate of KE and enstrophy transfers to the modal KE and modal enstrophy by nonlinearity; $\mathcal{F}_u({\bf k}), \mathcal{F}_\omega({\bf k})$ are respectively the modal KE and enstrophy injection rates by the external force; and $D_u ({\bf k}), D_\omega ({\bf k})$ are respectively the dissipation rates of the modal KE and enstrophy.  We define KE and enstrophy fluxes for a wavenumber sphere of radius $k_0$ as
\begin{eqnarray}
\Pi_u(k_0) & = & -\int_0^{k_0} T_u(k) dk, \\
\Pi_\omega(k_0) & = & -\int_0^{k_0} T_\omega(k) dk.
\end{eqnarray}

Using the above equations we can derive the following equations for one-dimensional spectra~\citep{Lesieur:book:Turbulence,Verma:book:BDF}
\begin{eqnarray}
 \frac{\partial}{\partial t}   E_u(k,t) & = & - \frac{\partial}{\partial k}  \Pi_u(k,t) + \mathcal{F}_u(k,t)  -D_u(k,t),  
 \label{eq:hydro_2d:Ek_energetics}  
 \end{eqnarray}
 \begin{eqnarray}
 \frac{\partial}{\partial t}   E_\omega(k,t) & = & - \frac{\partial}{\partial k}  \Pi_\omega(k,t) + \mathcal{F}_\omega(k,t)  -D_\omega(k,t).
 \label{eq:hydro_2d:Ew_energetics} 
\end{eqnarray}
In the next section we employ the above equations to derive spectra and fluxes for the KE and enstrophy in the inertial range and beyond.

\section{The spectra and fluxes of kinetic energy and enstrophy beyond the inertial range}
\label{sec:theory}

In this section we extend \citet{Kraichnan:PF1967_2D}'s formulas for the KE and enstrophy spectra and fluxes beyond the inertial range using Pao's  conjecture \cite{Pao:PF1965}.   For three-dimensional hydrodynamic turbulence, there are several models for the inertial-dissipation ranges.  They are by \citet{Pao:PF1965}, \citet{Pope:book}, and  \citet{Martinez:JPP1997}.  Among these models, one by  Pao provides the best fit to the numerical data, as demonstrated by \citet{Verma:FD2018}.   In addition, Pao's model has no additional parameter.  

Motivated by these successes, we attempted Pao's conjecture to model two-dimensional turbulence that has two regimes: $k < k_f$, and $k>k_f$.  Fortunately, the model predictions fit quite well with the numerical data inertial-dissipation range.  In literature it has been a challenge to model the $k>k_f$ range, but, as we demonstrate in the paper, Pao's model works quite well for this regime.

Under a steady state,   in Eqs.~(\ref{eq:hydro_2d:Ek_energetics}, \ref{eq:hydro_2d:Ew_energetics}), we set $\partial /\partial t =0$.  In addition, in the inertial range, the injection rates by the external force vanish, while the dissipation rates are negligible.  Hence,  
\be
\frac{d}{dk} \Pi_u(k) = 0;~~~  \frac{d}{dk} \Pi_\omega(k) =0
\label{eq:flux_const}
   \ee
   that leads to constancy of KE and enstrophy fluxes.   \citet{Kraichnan:PF1967_2D} showed that $\Pi_u(k)$ is constant for $k<k_f$, while $\Pi_\omega(k)$ is constant for $k>k_f$.  For these regimes, dimensional analysis yields~\citep{Kraichnan:PF1967_2D}
   \bea
 E_u(k) & = & C \bar{\Pi}_{u}^{2/3}k^{-5/3}~~~\mathrm{for}~k<k_f 
 \label{eq:E_u_5/3} \\
 E_u(k) & = & C' \bar{\Pi}_{\omega}^{2/3}k^{-3}~~~~\mathrm{for}~k>k_f 
  \label{eq:E_u_3}
 \eea
 where $\bar{\Pi}_{u}, \bar{\Pi}_{\omega}$ are respectively the values of the KE and enstrophy fluxes in the inertial range, and $C,C'$ are constants.   In Sec.~\ref{Sec:results} we show that $\bar{\Pi}_{u}, \bar{\Pi}_{\omega}$ differ from $\epsilon_{u}, \epsilon_{\omega}$ respectively.  In this paper we take the maximum value of the respective fluxes for $\bar{\Pi}_{u}, \bar{\Pi}_{\omega}$.
 
 To extend the above scaling beyond inertial range, but still away from the forcing range, we retain $D_u(k)$ and $D_\omega(k)$ in Eq.~(\ref{eq:flux_const}) that yields
 \bea
\frac{d}{dk} \Pi_u(k) & = &    -2 \nu k^2 E_u(k), \label{eq:hydro_2d:Ek_energetics_steady} \\
  \frac{d}{dk} \Pi_\omega(k)   & = &    -2 \nu k^2 E_\omega(k).
 \label{eq:hydro_2d:Ew_energetics_steady}  
   \eea
   The above relations are valid for all wavenumbers.   The above two equations have four unknowns, hence they cannot be uniquely solved.  To overcome this difficulty, we  extend Pao's conjecture~\citep{Pao:PF1968} for 3D hydrodynamic turbulence to 2D turbulence that enables us to extend the energy and enstrophy spectra beyond the inertial range.  In the following two sections we will describe them for $k<k_f$ and $k>k_f$ regimes separately.
  
 
\subsection{$k<k_f$}   
     We assume that  for $k<k_f$,  $E_u(k)/\Pi_u(k)$ is a function of $\bar{\Pi}_{u}$ and $k$, and it is independent of $\nu$ and the forcing parameters.  Under these assumptions, dimensional analysis yields
 \bea
 \frac{E_u(k)}{\Pi_u(k)} & = & -C \bar{\Pi}_{u}^{-1/3} k^{-5/3}.
 \label{eq:Eu_Pao_eqn}  
   \eea
   Note that the negative sign in Eq.~(\ref{eq:Eu_Pao_eqn}) is due to the fact that $\Pi_u(k) < 0$.  Substitution of Eq.~(\ref{eq:Eu_Pao_eqn})   in Eq.~(\ref{eq:hydro_2d:Ek_energetics_steady}) yields
    \bea
\frac{d}{dk} \Pi_u(k) & = &   2 C  \nu   \bar{\Pi}_{u}^{-1/3} k^{1/3} \Pi_u(k), \label{eq:hydro_2d:Pi_u_ode} 
   \eea
whose solution   is
  \begin{eqnarray}
\Pi_u(k)  & =  & -  \bar{\Pi}_u \exp{\left( \frac{3}{2} C  (k/k_d)^{4/3}\right)}, \label{eq:hydro_2d:Pao_Pik}\\
E_u(k)  & = & C   \bar{\Pi}_u^{2/3} k^{-5/3}  \exp{\left( \frac{3}{2} C  (k/k_d)^{4/3}\right)}, \label{eq:hydro_2d:Pao_Ek}
\end{eqnarray} 
where
$k_d =  \left({  \bar{\Pi}_u}/{\nu^3}\right)^{1/4}$, and $\bar{\Pi}_{u} > 0$. 

Now we investigate enstrophy flux in this regime.  It is generally conjectured that $\Pi_\omega(k) \approx 0$ in this regime~\citep{Boffetta:JFM2007, Boffetta:ARFM2012}.  But this is not the case because Eqs.~(\ref{eq:hydro_2d:Ek_energetics_steady}, \ref{eq:hydro_2d:Ew_energetics_steady}) yields
\be
\frac{d \Pi_\omega(k)}{d \Pi_u(k)} = k^2.
\ee
In fact, we can determine the enstrophy flux using Eqs.~(\ref{eq:hydro_2d:Ek_energetics_steady}) in the following manner.   Substitution of $E_u(k)$ of Eq.~(\ref{eq:hydro_2d:Pao_Ek}) in Eq.~(\ref{eq:hydro_2d:Ew_energetics_steady}) yields 
 \bea
 \Pi_\omega(k)   & = & -2 \nu   \int^k k'^4 E_u(k') dk' \nonumber \\
 & = &  -2 \nu C   \bar{\Pi}_u^{2/3} \int^k  k'^{7/3} \exp{\left( \frac{3}{2} C  (k'/k_d)^{4/3}\right)} dk' \nonumber \\
 & = &   -2 \nu C    \bar{\Pi}_u^{2/3} k_d^{10/3} 
 \int^x dx' x'^{7/3} \exp{\left( \frac{3}{2} C  x'^{4/3}\right)},
 \label{eq:hydro_2d:Piw_k<kf} \nonumber \\
 \eea
 where $x' = k/k_d$.  
 
 In the following subsection we will describe the energy flux and spectrum, as well as enstrophy flux in the $k>k_f$ regime.

\subsection{$k>k_f$}
 In this regime we assume that  $E_\omega(k)/\Pi_\omega(k)$  is a function of $\bar{\Pi}_\omega$ and $k$, and it is independent of $\nu$ and forcing function.  This assumption leads to 
  \bea
   \frac{E_\omega(k)}{\Pi_\omega(k)} & = & C'  \bar{\Pi}_{\omega}^{-1/3} {k^{-1}}.  \label{eq:Ew_Pao_eqn} 
   \eea
    Substitution of Eqs.~(\ref{eq:Ew_Pao_eqn})   in Eqs.~(\ref{eq:hydro_2d:Ew_energetics_steady})  yields
    \bea
  \frac{d}{dk} \Pi_\omega(k)   & = &    -2 C'  \nu  \bar{\Pi}_{\omega}^{-1/3}  k^{-1} \Pi_\omega(k),
 \label{eq:hydro_2d:Pi_w_ode}  
   \eea
whose solution  is
\begin{eqnarray}
\Pi_{\omega}(k) & =  &  \bar{\Pi}_\omega \exp{\left(- C'    (k/k_{d2D})^{2}\right)}, \label{eq:hydro_2d:Piwk_omega_inertial_diss} \\
E_\omega(k) & = & C'   \bar{\Pi}_\omega^{2/3} k^{-1} \exp{\left(-  C' (k/k_{d2D})^{2}\right)}, \label{eq:hydro_2d:Ek_omega_inertial_diss}  \\
  E_u(k) &= & C'   \bar{\Pi}_\omega^{2/3} k^{-3} \exp{\left(-  C'    (k/k_{d2D})^{2}\right)},   \label{eq:hydro_2d:Eu_k>k_f}
\end{eqnarray}
where $k_{d2D} = { \bar{\Pi}_\omega^{1/6}}/{\sqrt{\nu}}$
is the enstrophy dissipation wavenumber.   Note that $  E_u(k) $ of Eq.~(\ref{eq:hydro_2d:Eu_k>k_f}) is steeper than $k^{-3}$ in the inertial-dissipation range.  The strong gaussian factor dominates $k^{-3}$ scaling; this could be reason for the difficulty in observing $k^{-3}$ spectrum in $k>k_f$ regime.  In Sec.~\ref{Sec:results} we show consistency of the above steepening with the numerical results.

To determine $\Pi_u(k)$, we substitute $E_u(k)$ of Eq.~(\ref{eq:hydro_2d:Eu_k>k_f}) in Eq.~(\ref{eq:hydro_2d:Ek_energetics_steady}), and integrate the equation from $k$ to $\infty$.  Using $\Pi_u(\infty)=0$ and making a change of variable $x=C(k/k_{d2D})^2$, we obtain
  \begin{eqnarray}
\Pi_u(k) & = &  2 \nu   \int_{k}^\infty k'^2 E_u(k') dk' \nonumber \\
& = &  \frac{ \bar{\Pi}_\omega {C'}}{  k_{d2D}^2} \int_{C(k/k_{k2dD})^2}^\infty \frac{1}{x} \exp{(- x)} dx \nonumber \\
&=  & \frac{ \bar{\Pi}_\omega {C'}}{  k_{d2D}^2}  E_1(C' (k/k_{d2D})^2),
\label{eq:hydro_2d:Piuk>kf}
\end{eqnarray}
where $E_1$ is the exponential integral~\cite{Abramowitz:book}. 

Since $k_{d2D}$ represents the enstrophy dissipation wavenumber, hence $k_f/k_{d2D}<  k/k_{d2D} \lessapprox 1$.  In this range, $E_1(x) $ is of the order of unity.  For example, $E_1(x) < 2$ for $0.1 < x < 1.6$~\cite{Abramowitz:book}.  Hence, using Eq.~(\ref{eq:hydro_2d:Piuk>kf}) and the fact that $ k_{d2D} \gg 1$, we deduce that
\be
\Pi_u(k) \approx \frac{ \bar{\Pi}_\omega C'}{k_{d2D}^2} \rightarrow 0
\label{eq:hydro_2d:Piu_approx_k>kf}
\ee
That is, $\Pi_u(k) \ll \bar{\Pi}_\omega$ in the forward enstrophy regime. This observation is consistent with the findings of \citet{Gotoh:PRE1998} that $\Pi_u(k) \rightarrow 0$ in the inertial-dissipation range ($k > k_f$).  However, the functional form of $\Pi_u(k) $ in \citet{Gotoh:PRE1998}  differs from that of Eq.~(\ref{eq:hydro_2d:Piuk>kf}).

Note however that the aforementioned model of spectra and fluxes of KE and enstrophy assume steady state.  As we show in Sec.~\ref{Sec:results}, this assumption does not hold due to unsteady nature of 2D turbulence.  It has been reported that the large-scale KE grows with time.  As a result, some of the above predictions match with the simulation results, while some do not.
   
We will attempt to verify the above scaling functions using numerical simulations.

\section{Details of Numerical Simulations}
\label{Sec:simdetails}

\begin{table*}[ht]
\caption{ Table containing total and partial viscous dissipation rates (columns 2, 3), and total and partial enstrophy dissipation (columns 5, 6). Here, $\mathcal{M} \mathrm{(|\Pi_{u}(k<k_f)|)}$ and $\mathcal{M}\mathrm{(|\Pi_{\omega}(k>k_f)|)}$ represent $\mathrm{max(|\Pi_{u}(k<k_f)|)}$ and $\mathrm{max(|\Pi_{\omega}(k>k_f)|)}$ respectively. }
 \label{tab:epsilon}
\centering
\begin{tabular}{ | P{1.6cm} | P{2.8cm}| P{2.5cm} | P{2.1cm} | P{2.7cm} | P{2.8cm} | P{2.1cm} | }
\hline
Data type & \scriptsize{$ 2 \nu   \int_0^{k_\mathrm{max}} k'^2 E_u(k') dk' $}  & \scriptsize{$2 \nu   \int_0^{k_f} k'^2 E_u(k') dk' $} & \scriptsize{$\mathcal{M}{(|\Pi_{u}(k<k_f)|)}$} &  \scriptsize{$ 2 \nu   \int_0^{k_\mathrm{max}}  k'^4 E_u(k') dk' $} &  \scriptsize{$2 \nu   \int_{k_f}^{k_\mathrm{max}} k'^4 E_u(k') dk' $} &   \scriptsize{$\mathcal{M}\mathrm{(|\Pi_{\omega}(k>k_f)|)}$} \\
 \hline
 Single time frame data of $2048^{2}$ grids & 24.0 & 16.4 &  16.1 & $5.54 \times 10^{4}$ & $3.64 \times 10^{4}$ &   $3.24 \times 10^{3}$    \\
  \hline
 Single time frame data of $8192^{2}$ grids & 18.8 & 17.5  & 77.9 &  $8.07 \times 10^{4}$ & $2.78 \times 10^{4}$ &  $1.10 \times 10^{4}$  \\
 \hline
 Time averaged data of $2048^{2}$ grids & 21.8 & 11.8    & 4.73 & $5.24 \times 10^{4}$ & $2.73 \times 10^{4}$   & $4.00 \times 10^{3}$ \\
 \hline
 Time averaged data of $8192^{2}$ grids & 15.1 & 14.2  & 43.4 &  $5.94 \times 10^{4}$ & $2.12 \times 10^{4}$ & $1.21 \times 10^{4}$ \\
 \hline
\end{tabular}
 \end{table*}

\begin{table}
\centering
\caption{ Table  listing $k_d$ and $k_{d2D}$ for the four cases given in Table~\ref{tab:epsilon}.}
 \label{tab:kd}
\begin{tabular}{ |c | c| c|  } 
\hline
Data type &  \ \ \ \scriptsize{$k_{d}$}  & \ \ \ \scriptsize{$k_{d2D}$} \\
 \hline
 Single time frame data of $2048^{2}$ grids   & $3.56\times 10^{2}$ &   $2.94 \times 10^{2}$   \\
  \hline
 Single time frame data of $8192^{2}$ grids   & $1.30\times 10^{3}$ &   $1.00\times 10^{3}$  \\
 \hline
 Time averaged data of $2048^{2}$ grids  &    $2.62 \times 10^{2}$ & $3.15\times 10^{2}$ \\
 \hline
 Time averaged data of $8192^{2}$ grids     &  $1.13 \times 10^{3}$ & $9.46\times 10^{2}$  \\
 \hline
\end{tabular}

 \end{table}

 In the  present paper, we perform numerical simulations of forced 2D hydrodynamic turbulence using spectral method. The system is doubly-periodic in a domain of size $2\pi \times 2\pi$.  We employ two different grid resolutions $2048^{2}$ and $8192^2$ to make sure that our results are grid independent. 
 The equations are solved using a fully dealiased, parallel pseudo-spectral code TARANG~\cite{Chatterjee:JPDC2018} with fourth-order Runge-Kutta time marching scheme. For dealiasing purpose, 2/3-rule is chosen~\cite{Canuto:book:SpectralFluid, Boyd:book:Spectral}. The viscosity for the $2048^{2}$ and $8192^2$ grids are  set at $1\times 10^{-3}$ and $3\times 10^{-4}$ respectively that yields  corresponding Reynolds numbers of $1.2 \times 10^{4}$ and $4.2 \times 10^{5}$.   For our simulation we employ eddy turnover as unit of time.   The simulations for the two grids were run up to $t_\mathrm{final} = 10.0$ and $1.74$ respectively.  We employ Courant-Friedrichs-Lewy  (CFL) condition to determine the timestep $dt$.  For the two grids,  the average time steps are $3.4 \times 10^{-5}$ and $6.9 \times 10^{-5}$ respectively.
 
We force the flow at wavenumber band $k_f = (50, 51)$ and (100, 101)  for $2048^{2}$ and $8192^2$ grids respectively.  These resolutions provide more than a decade of inverse cascade regime.  The enstrophy cascades forward in  $k>k_f$ regime, but the spectrum is steeper than $k^{-3}$ due to the dissipation effects.  

Using the numerical data we compute the one-dimensional energy and enstrophy spectrum using 
\bea
E_u(k) & = & \frac{1}{2} \sum_{k-1 < |{\bf k'}|\le k} |{\bf u(k')}|^2, \\
E_\omega(k) & = & \frac{1}{2} \sum_{k-1 < |{\bf k'}| \le k} |\omega {\bf(k')}|^2.
\eea
The energy flux $\Pi_u(k_0)$ is defined as the energy leaving the sphere of radius $k_0$ due to nonlinearity.  This quantity is computed as~\citep{Dar:PD2001,Verma:PR2004}
\be
 \Pi_u(k_0) = \sum_{|{\bf p}| \le k_0} \sum_{|{\bf k }| > k_{0}} S^{uu}({\bf k|p|q})
  \label{eq:Pi_u}
 \ee
 where
 \begin{eqnarray}
S^{uu}({\bf k|p|q}) &=& \Im \left[ {\bf  \{  k \cdot u(q) \} \{ u(p) \cdot u^{*}(k) \} } \right],
\label{eq:suu} 
\end{eqnarray} 
is the  {\em mode-to-mode} energy transfer  from Fourier mode ${\bf u(p)}$ to Fourier mode ${\bf u(k)}$ with Fourier mode ${\bf u(q)}$ acting as a mediator.  Note that the wavenumbers ${\bf (k,p,q)}$ form a triad with ${\bf k=p+q}$.   Similarly, we compute the enstrophy flux as
\be
 \Pi_\omega(k_0) = \sum_{|{\bf p}| \le k_0} \sum_{|{\bf k }| > k_{0}} S^{\omega \omega}({\bf k|p|q})
 \label{eq:Pi_w}
 \ee
 where
 \begin{eqnarray}
S^{\omega \omega}({\bf k|p|q}) &=& \Im \left[ {\bf  \{  k \cdot u(q) \} \{ \omega(p)  \omega^{*}(k) \} } \right],
\label{eq:sww} 
\end{eqnarray} 
is the  {\em mode-to-mode} enstrophy transfer  from Fourier mode ${\bf \omega(p)}$ to Fourier mode ${\bf \omega(k)}$ with Fourier mode ${\bf u(q)}$ acting as a mediator.  

The energy and enstrophy fluxes provide insights into the global  transfers in the system.  For a more detailed picture, we compute the shell-to-shell energy transfers.  We divide the wavenumber space into various concentric shells.  The shell-to-shell energy transfer from shell $m$ to shell $n$ is given by 
\begin{equation}
T^{u,m}_{u,n}=\sum_{\mathbf{p}\in m} \sum_{\mathbf{k}\in n}S^{uu}(\mathbf{k|p|q}).
\label{eq:S2S}
\end{equation}
In the present work, we compute the shell-to-shell energy transfer for $k<k_f$ regime.  We compare  our numerical result with those for 3D hydrodynamic turbulence for which the energy transfer is local and forward in the inertial range, that is, the dominant energy transfer is  from shell $m$ to $m+1$.   For better resolution, we perform the shell-to-shell transfer computations for $8192^2$ grid with log-binned shells.   We divide the Fourier space into 20 concentric shells; the inner and outer radii of the $i^{th}$ shell are $k_{i-1}$ and $k_{i}$ respectively. The shell radii for $N=8192^{2}$ grids are  $k_{i}=\lbrace0, 2, 4, 8,  8\times2^{s(i-3)}, ..., 2048, 4096\rbrace$, where $s=8/15$ and $i$ is the shell index. Inertial range shells have been chosen by
logarithmic binning  because of the power law physics in the inertial range.

In this paper we do not report the shell-to-shell enstrophy transfer in the $k>k_f$ regime due to lack of constant enstrophy regime because of strong dissipation.  In future we plan to perform simulations with $k_f \approx 1$ that would provide significant wavenumber range  of constant enstrophy flux.  It will be meaningful to perform shell-to-shell enstrophy transfer computations using such data.

 As we will describe in the following section, the energy and enstrophy fluxes in the $k<k_f$ are highly fluctuating, possibly due to unstable nature of 2D turbulence.  Therefore, in addition to illustrating the above fluxes for a single snapshot for $2048^2$ and $8192^2$ grids, we also present averages of these fluxes over {\color{blue}35} frames.
 
 The formulas for the energy and enstrophy fluxes and spectra described in previous section requires values of $C, C',  \bar{\Pi}_u,  \bar{\Pi}_\omega, k_d$, and $k_{d2D}$.  Numerical simulations and analytical calculations~\cite{Smith:PRL1993,Gotoh:PRE1998,Boffetta:ARFM2012,Nandy:IJMPB1995}  predict that $C \approx 6.5 \pm 1$, and $C' \approx 1$ to 2.    In this paper we choose $C=6.5$ and $C'=1.0$ consistent with the above results.  The estimation of the dissipation rates $ \bar{\Pi}_u,  \bar{\Pi}_\omega$  for two-dimensional turbulence is tricky.  Since the flow is forced at intermediate wavenumber band, we compute the energy and enstrophy dissipation rates in both, $k < k_f$ and $k>k_f$, regimes and list them in Table~\ref{tab:epsilon}.  As shown by the entries of the table, there are strong energy and enstrophy dissipation in $k < k_f$ band, contrary to three-dimensional hydrodynamic turbulence where dissipation occurs at large $k$'s.  This is because of the large magnitude of $E_u(k)$ in this regime.  Also, for the total dissipation rates, $\epsilon_\omega \approx k_f^2 \epsilon_u$, as expected.  Interestingly, none of these dissipation rates match with maximum values of the energy and enstrophy fluxes, which are denoted by $\mathcal{M}(|\Pi_u|)$ and $\mathcal{M}(|\Pi_\omega|)$ respectively.  For the best fit to the numerical results of Sec.~\ref{Sec:results}, we take $\bar{\Pi}_{u} = \mathcal{M}{(|\Pi_{u}(k<k_f)|)}$ and $\bar{\Pi}_{\omega} = \mathcal{M}{(|\Pi_{\omega}(k>k_f)|)}$.   

Lastly, we estimate  $k_{d}=(\epsilon_{u}/\nu^{3})^{1/4}$.  However, the forward enstrophy cascade regime does not have a significant $k^{-3}$ power law regime for the spectrum, hence we cannot use the formula {\color{blue}$k_{d2D} = {\epsilon_\omega^{1/6}}/{\sqrt{\nu}}$} for its estimation.  Rather, we obtain $k_{d2D} $ from the best fit curve to the enstrophy flux.  These parameters are listed in Table~\ref{tab:kd}.  We employ these parameters in Eqs.~(\ref{eq:hydro_2d:Pao_Pik}, \ref{eq:hydro_2d:Pao_Ek}, \ref{eq:hydro_2d:Piwk_omega_inertial_diss}, \ref{eq:hydro_2d:Eu_k>k_f}, \ref{eq:hydro_2d:Piuk>kf})  to compute the best fit curves for modelling the numerical results.

In the next section we will report numerical results on the spectra and fluxes of energy and enstrophy.

\section{Results and discussions} 
\label{Sec:results}

\begin{figure}[htbp]
\begin{center}
\includegraphics[scale = 0.8]{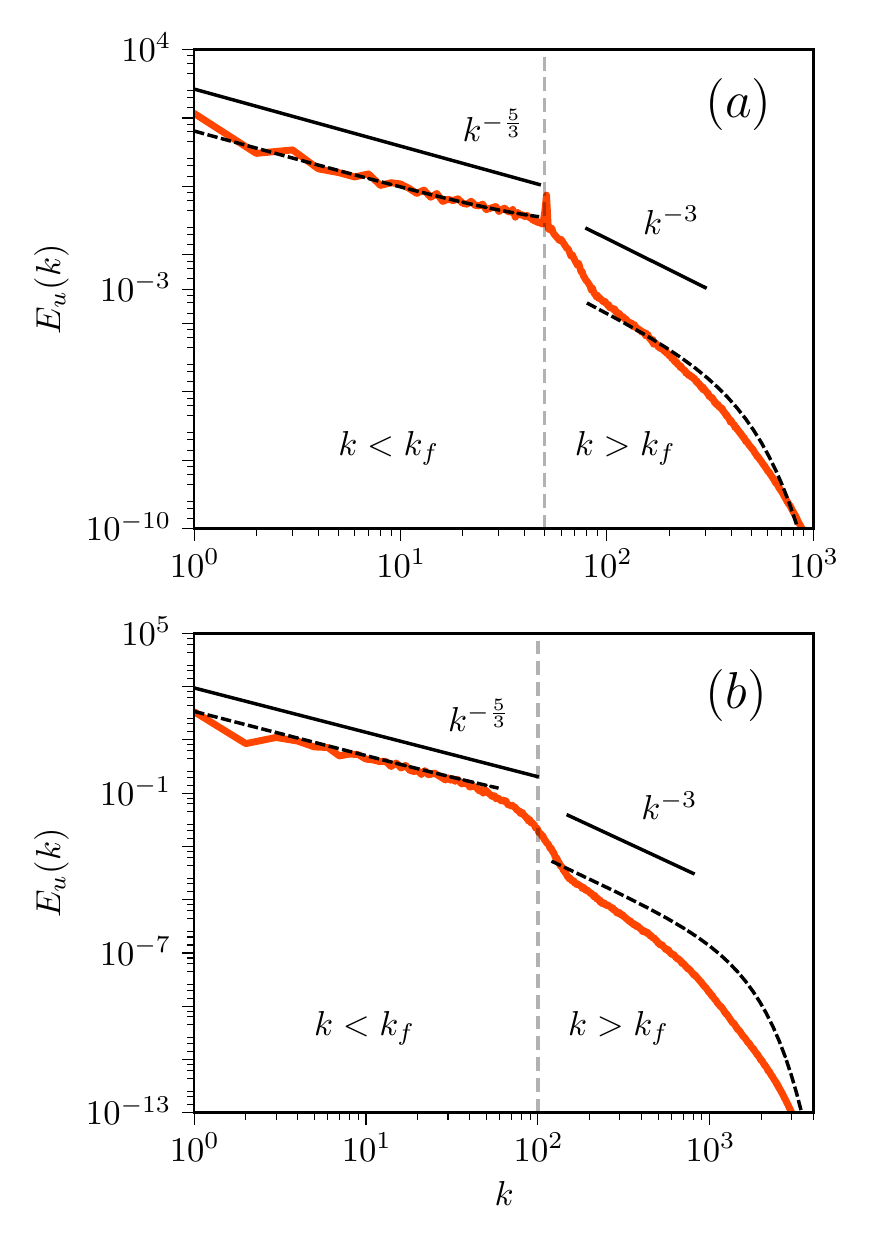}
\end{center}
\caption{Kinetic energy spectrum for a single time frame  of 2D forced turbulence: (a) using $2048^2$ grid simulation with forcing at  $k_f=(50,51)$, (b) using $8192^2$ grid simulation with forcing at $k_f=(100,101)$. The forcing wavenumbers are indicated by dashed vertical lines. The plots exhibit numerical results (solid red curves), model predictions [Eqs.~(\ref{eq:hydro_2d:Pao_Ek}, \ref{eq:hydro_2d:Eu_k>k_f})] (dashed black curves) and   \citet{Kraichnan:PF1967_2D}'s predictions (solid black lines).   The model predictions match with the numerical results quite well.  Fitting parameters are given in Tables.~\ref{tab:epsilon} and \ref{tab:kd}.} 
\label{fig:hydro_2d:Eu}
\end{figure}

In the present section we will report the numerically computed spectra and fluxes of energy and enstrophy.  We will compare these results with the model predictions of Sec.~\ref{sec:theory}.

\subsection{Energy spectra}
In this subsection, we describe  the energy and enstrophy spectra of  2D turbulence. In Fig.~\ref{fig:hydro_2d:Eu}(a,b) we plot these spectra for $2048^{2}$ and $8192^{2}$ grids.  The numerically computed  spectra are exhibited using red  solid curves, and the model predictions  of Sec.~\ref{sec:theory} using dashed  black curves.  In this figure we also plot the predictions of Kriachnan's theory using solid black lines.    We observe that the model  predictions match with the numerical results reasonably well.

For $k < k_f$, the numerical results and model predictions yield $E_u(k) \sim k^{-5/3}$, which is the prediction of \citet{Kraichnan:PF1967_2D}. The curves tend to increase relative to $k^{-5/3}$, though very slowly, due to the exponential factor   of Eq.~(\ref{eq:hydro_2d:Pao_Ek}).   For $k>k_f$, $E_u(k)$ is steeper than  \citet{Kraichnan:PF1967_2D}'s predictions of $k^{-3}$\citep{Legras:1988,Kellay:PRL1995,Gotoh:PRE1998}. The steepening of $E_u$ compared to $k^{-3}$ is due to the dissipative effects, more so because of the $\exp(-(k/k_{d2D})^2)$ factor.  The model equation~(\ref{eq:hydro_2d:Eu_k>k_f})  overestimates the energy spectrum.  This is possibly due to the lack of inertial range, and due to uncertainties in $k_{d2D}$.    A more refined simulation is required to decipher this issue.   Also, we compared our numerical results and predictions with those of \citet{Gotoh:PRE1998} in the limiting cases, and observed general agreement.    Note that  $E_\omega(k) = k^2E_u(k)$, hence $E_\omega(k)$ is not reported separately.

In the next subsection we will describe the energy and enstrophy fluxes computed  using the numerical data, as well as those predicted by the model of Sec.~\ref{sec:theory}.

\subsection{Energy and enstrophy fluxes}
\label{sec:flux}

\begin{figure*}[htbp]
\begin{center}
\includegraphics[scale = 0.7]{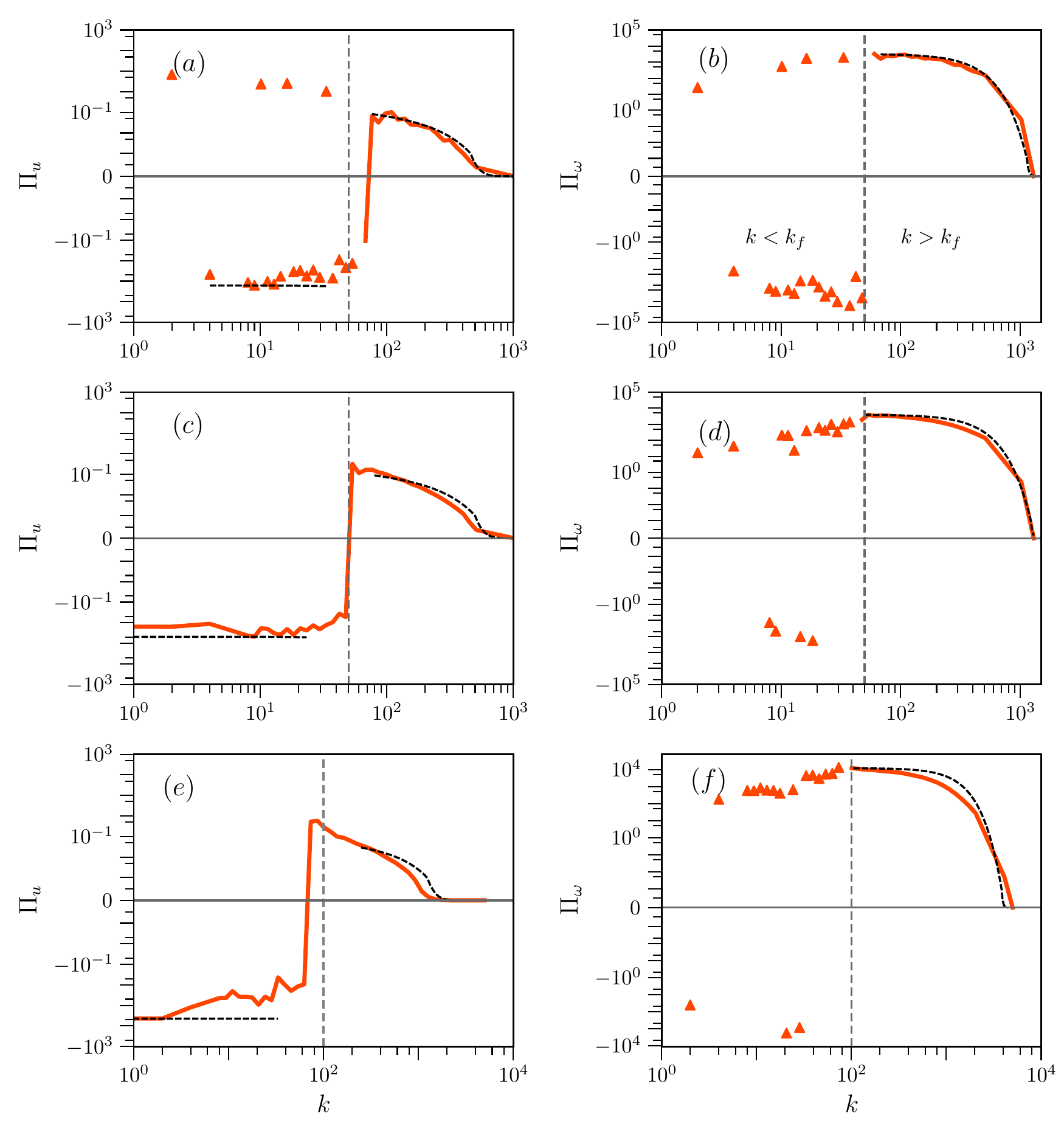}
\end{center}
\caption{Plots of kinetic energy flux, $\Pi_u(k)$ (left panel), and enstrophy fluxes, $\Pi_\omega(k)$ (right panel): (a,b) For a single time frame of $2048^2$ run; (c,d) Time averaged $\Pi_u(k)$ and  $\Pi_\omega(k)$ over  35 frames of  $2048^{2}$ run; (e, f) Time averaged $\Pi_u(k)$ and  $\Pi_\omega(k)$  over 35 frames of  $8192^{2}$ run.  The numerical results (red solid lines) match with the model predictions [Eqs.~(\ref{eq:hydro_2d:Piwk_omega_inertial_diss}, \ref{eq:hydro_2d:Piuk>kf})] (black dashed lines) in the $k>k_f$ regime. The discrepancies in the $k<k_f$ regime are possibly due to the unsteady nature of the flow.    Note that the plots are in loglog scale because of huge range of scales of  $\Pi_u(k)$ and  $\Pi_\omega(k)$.}
\label{fig:hydro_2d:flux_Piu1}
\end{figure*}

We  compute  the energy and enstrophy fluxes for 2D  turbulence using Eqs.~(\ref{eq:Pi_u}, \ref{eq:Pi_w}).  We compute these quantities for both the grid resolutions ($2048^2$ and $8192^2$).   The numerically computed fluxes are exhibited in Fig.~\ref{fig:hydro_2d:flux_Piu1}.   The left and right panels of Fig.~\ref{fig:hydro_2d:flux_Piu1}  illustrate the energy and enstrophy fluxes respectively.  

\begin{figure}[htb]
\begin{center}
\leavevmode
\includegraphics[scale = 0.8]{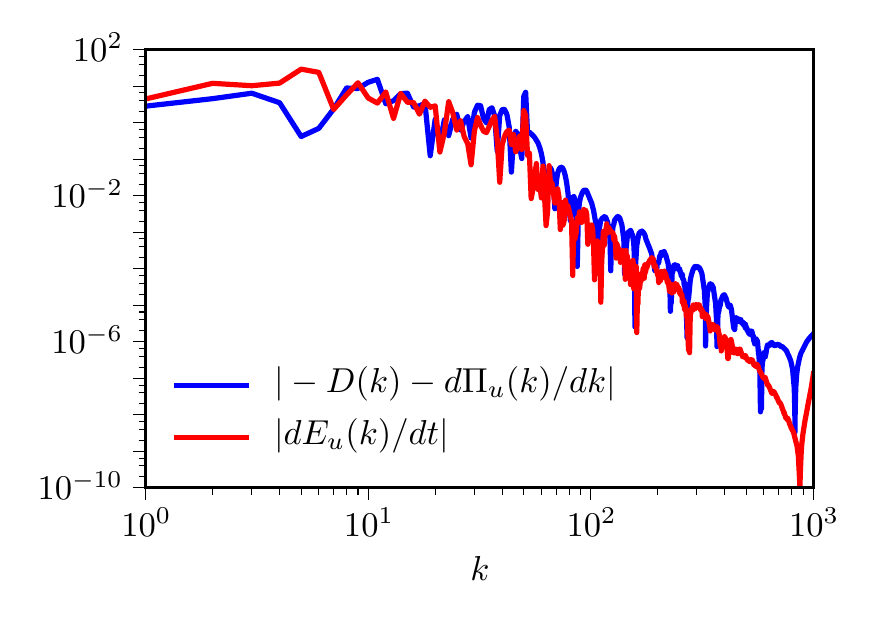}
\end{center}
\caption{For the $2048^{2}$ grid simulation, plots of $|-D(k)-d\Pi_{u}(k)/dk|$ and $|\partial{E(k,t)}/\partial{t}|$. Significant measures of $|\partial{E(k,t)}/\partial{t}|$ indicates unsteady nature of the flow.  }
\label{fig:hydro_2d:energyrate1}
\end{figure}

 \begin{figure*}[htbp]
\begin{center}
\includegraphics[scale = 0.75]{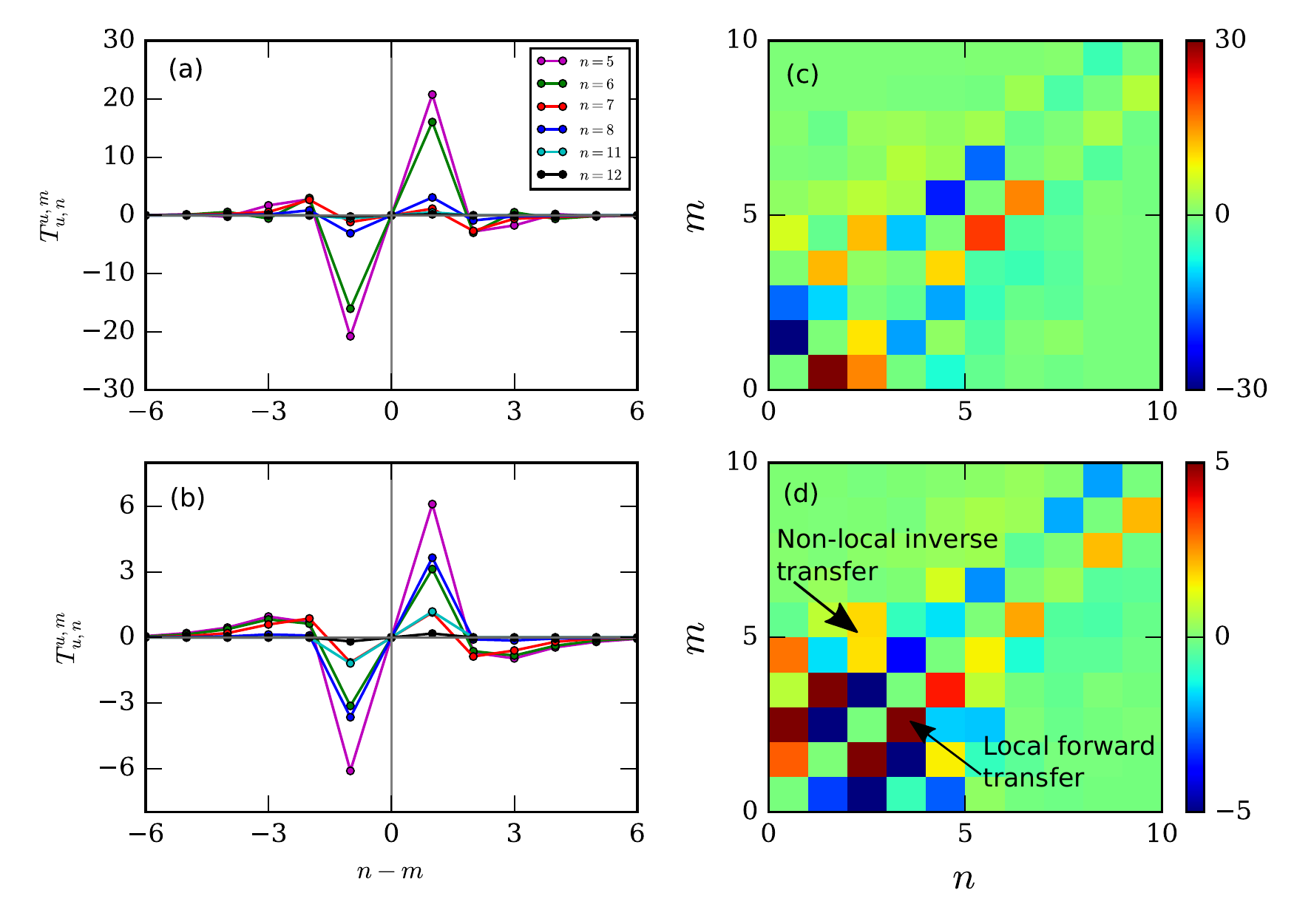}
\end{center}
\caption{Plots of shell-to-shell energy transfers $T^{u,m}_{u,n}$ vs.~$n-m$ for the $8192^2$ grid run in the inertial range shells of $k<k_f$ regime:  (a) For a single time frame; (b) For time-averaged data with 35 frames.  (c,d) Density plots of shell-to-shell energy transfers $T^{u,m}_{u,n}$ corresponding to (a,b) respectively.  The $x,y$ axes represent the receiver and giver shells respectively. The energy transfers are local and forward for neighbouring shells, but nonlocal and backward for distant shells. }
\label{fig:hydro_2d:Tnm_n_minus_m0}
\end{figure*}

The top panel of Fig.~\ref{fig:hydro_2d:flux_Piu1} exhibits the energy and enstrophy fluxes for a single time frame of $2048^2$ grid run.  We observe that these fluxes exhibit significant fluctuations for $k<k_f$.  Therefore, we compute average fluxes.  The fluxes in the middle and bottom panels are computed by averaging over  {\color{blue}35}  different time frames for $2048^2$ and $8192^2$ grids respectively.  In the plots the red curves represent the numerically computed fluxes, while the black dashed lines represent the model predictions of Sec.~\ref{sec:theory}.

A careful observation of the figure shows that for $k > k_f$, the model predictions of the KE and enstrophy fluxes, Eqs.~(\ref{eq:hydro_2d:Piwk_omega_inertial_diss}, \ref{eq:hydro_2d:Piuk>kf}), are in good agreement with the numerical results. Thus,  Eqs.~(\ref{eq:hydro_2d:Piwk_omega_inertial_diss}, \ref{eq:hydro_2d:Piuk>kf}) describe 2D turbulence satisfactorily for the $k > k_f$ regime.    Note that in this regime, both $\Pi_u(k)$ and $\Pi_\omega(k)$ fall very sharply due to the gaussian nature of the exponential factor ($\exp(-(k/k_{d2D})^2)$).  Also, $\Pi_u(k) \ll \Pi_\omega(k)$, consistent with Eq.~(\ref{eq:hydro_2d:Piu_approx_k>kf}) and the predictions of \citet{Gotoh:PRE1998}.   This is somewhat surprising  that the model predictions overestimate the energy spectrum in this regime. This issue needs a further investigation.

However,  for $k < k_f$, the model predictions fail to describe  the numerical results well.  As shown in Fig.~\ref{fig:hydro_2d:flux_Piu1}(a,b), the energy and enstrophy fluxes computed using a single frame data exhibits significant fluctuations.   The fluctuations are somewhat suppressed on averaging, as shown in Fig.~\ref{fig:hydro_2d:flux_Piu1}(c-f), yet the enstrophy flux for $8192^2$ grid shows large fluctuations in $k<k_f$ regime.   We believe that the fluctuations in the fluxes are due to the unsteady nature of the flow.

Two-dimensional turbulence exhibits  inverse cascade of kinetic energy that leads to  formation of large-scale structures.  An imbalance between the viscous dissipation and the energy feed at the large-scale structures makes the flow unsteady.   As a result, Eqs.~(\ref{eq:hydro_2d:Ek_energetics_steady}, \ref{eq:hydro_2d:Ew_energetics_steady}) are not valid for $k<k_f$ regime.   To quantify the unsteadiness of the flow,  we  compute all the terms of Eq.~(\ref{eq:hydro_2d:Ek_energetics}) for $2048^2$ grid and
compare them. In Fig.~\ref{fig:hydro_2d:energyrate1}
 we plot $|\partial{E(k)}/\partial t|$ and $|-d\Pi_{u}(k)/dk-D_{u}(k)|$ with respect to $k$.   Though the left-hand and right-hand sides of Eq.(\ref{eq:hydro_2d:Ek_energetics}) match with each other,
noticeably, $|\partial{E(k)}/\partial t|$ is significant for $k<k_f$.  Note however that these quantities are small for $k>k_f$.  This is the reason for the  unsteady nature of the flow that leads to strong fluctuations in $\Pi_{u}(k)$ and $\Pi_{\omega}(k)$ in the $k<k_f$ regime.

 In the next subsection, we describe the shell-to-shell energy transfers for 2D turbulence.

\subsection{Shell to shell energy transfers}
In the present subsection we describe  the shell-to-shell energy transfers  for 2D  turbulence.  We compare our results with  three-dimensional turbulence for which the shell-to-shell  transfers are local and forward~\cite{Domaradzki:PF1990,Verma:Pramana2005S2S}.   

 We compute the shell-to-shell energy transfers in the wavenumber band $k<k_f$ using the formula of Eq.~(\ref{eq:S2S}).  As described in Sec.~\ref{Sec:simdetails}, for $8192^2$ grid simulation we divide this wavenumber region into $20$ shells.  The computed transfers are exhibited in Figures~\ref{fig:hydro_2d:Tnm_n_minus_m0}.

 In Fig.~\ref{fig:hydro_2d:Tnm_n_minus_m0}(a), we plot the shell-to-shell  energy transfers   $T^{u,m}_{u,n}$ vs. $n-m$  computed for a single frame.  Here $m,n$ are the giver and receiver shells respectively.  These transfers exhibit significant fluctuations for different data sets, hence we average the transfer rates for 35 frames.   The averaged transfers are exhibited in Fig.~\ref{fig:hydro_2d:Tnm_n_minus_m0}(b). As shown in the figures, specially Fig.~\ref{fig:hydro_2d:Tnm_n_minus_m0}(b),  shell $n$ receives energy from shell $n-1$ and  gives energy to shell $n+1$. Hence, among the nearest neighbour shells in the inertial range of $k<k_f$, the energy transfer in 2D hydrodynamic turbulence is forward.  Note however that $T^{u,m}_{u,n} < 0$ when $n-m > 2$ or 3 (for some shell).  This implies that shell $n$ receives energy  from far away shells. Therefore, in 2D hydrodynamic turbulence, the shell-to-shell energy transfers to the neighboring shells are forward, but they are backward for the distant shells.  
 
 In Fig.~\ref{fig:hydro_2d:Tnm_n_minus_m0}(c,d), we exhibit the corresponding density plots.   Here the indices of the $x$, $y$ axes represent the receiver and giver shells respectively.  Though the plot exhibit significant fluctuations,  the density plots are consistent with the results of Fig.~\ref{fig:hydro_2d:Tnm_n_minus_m0} (a,b).
 
The aforementioned shell-to-shell energy transfers of 2D turbulence differ significantly from the those of  3D turbulence for which the transfers are local and forward for the inertial range shells.  This divergence between the 2D and 3D flows is due to the inverse cascade of energy.  \citet{Verma:Pramana2005S2S} computed the shell-to-shell energy transfers for 2D turbulence using field-theoretic tools and reported local forward and nonlocal backward energy transfers for $k<k_f$ (consistent with the aforementioned numerical simulations); they showed that these complex transfers add up to yield a negative $\Pi_u(k)$.  Here, the nonlocal backward energy transfers from many shells play a critical role. 


We summarise our results in the next section. 

\section{Conclusions}
\label{Sec:conclusion}

In this paper we present several results on 2D forced turbulence with forcing employed at intermediate scales. Using Pao's conjecture, we extend \citet{Kraichnan:PF1967_2D}'s power law predictions for 2D turbulence beyond the inertial range.  In the new scaling solution, the power laws are coupled with exponential functions of $k$. 

To test the model predictions, we performed numerical solution of 2D turbulence on $2048^{2}$ and $8192^{2}$ grids with forcing at (50,51) and (100,101) wavenumber bands respectively.  We computed the spectra and fluxes of energy and enstrophy using the numerical data, and compared them with the model predictions.  We observe that the model predictions and numerical results on the spectra and fluxes of energy and enstrophy agree with each other for $k>k_f$.  In this regime, the energy spectrum is steeper than $k^{-3}$, which is primarily due to the exponential factor  $\exp(-(k/k_{d2D})^{2} )$ of Eq.~(\ref{eq:hydro_2d:Eu_k>k_f}).   The situation however is different for $k<k_f$.  Though the energy spectrum follows $k^{-5/3}$ power law, the energy and enstrophy fluxes exhibit significant fluctuations.  We show that these fluctuations arise due to the unsteady nature of the flow and inverse energy cascade of kinetic energy.  The fluctuations are somewhat suppressed  on averaging.  These issues need further investigation.

We also compute the shell-to-shell energy transfers in the $k<k_f$ regime.  We observe forward energy transfers for the nearest neighbour shells, but backward energy transfers for the other shells, consistent with the analytical findings of \citet{Verma:Pramana2005S2S}.  The nonlocal backward transfers add up to yield a negative energy flux.   In addition, we observe that the shell-to-shell energy transfers in $k<k_f$ regime exhibits significant fluctuations among different frames due to the unsteady nature of the flow.

We also remark that rapidly rotating turbulence, and magnetohydrodynamic and quasi-static magnetohydrodynamic turbulence with strong external magnetic field exhibits quasi 2D behaviour~\cite{Oks:POF2017,Xia:POF2017,Sharma:PF2018,Potherat:JFM2000,Lee:ApJ2003,Reddy:PF2014,Verma:ROPP2017}.  \citet{Sharma:PF2018} employed the enstrophy flux derived in Sec.~\ref{sec:theory} to describe the energy spectra and flux of rapidly rotating turbulence.  Similar attempts  have been made to explain the turbulence properties of quasi-static magnetohydrodynamic turbulence~\cite{Potherat:JFM2000,Lee:ApJ2003,Reddy:PF2014,Verma:ROPP2017}.  In addition, two-dimensional magnetohydrodynamic turbulence too exhibits interesting properties~(e.g., see \cite{Mininni:PF2005}), but these discussions are beyond the scope of this paper.

In summary, our findings on 2D turbulence sheds interesting light on the energy and enstrophy transfers.  In future, we plan to extend the present work to the extended regime of constant enstrophy flux, in particular study the shell-to-shell enstrophy transfers and explore whether they are local or nonlocal. 
 
\begin{acknowledgments}
We thank Manohar Sharma for useful discussions and Shaswant Bhattacharya for comments on the manuscript. Our numerical simulations were performed on Shaheen II at {\sc Kaust} supercomputing laboratory, Saudi Arabia, under the project k1052.  This work was supported by the research grants PLANEX/PHY/2015239 from Indian Space Research Organisation, India, and by the Department of Science and Technology, India (INT/RUS/RSF/P-03) under the Indo-Russian project, and IITK institute postdoctoral fellowship.   
\end{acknowledgments}


%

\end{document}